\documentclass[amsmath, amssymb,10pt,aps,prb,twocolumn,notitlepage,showpacs,superscriptaddress]{revtex4-1}
\usepackage{graphicx}
\usepackage{calc}
\usepackage{braket}
\usepackage{ulem}   
\usepackage{slashed}
\usepackage{wasysym}
\usepackage{siunitx}
\usepackage{dsfont}
\usepackage{amsthm,amsmath,amsfonts,amssymb,verbatim,color}
\usepackage{graphicx}
\usepackage{subfigure}
\usepackage{bm}
\usepackage{epsfig,slashed}
\usepackage[T1]{fontenc}
\usepackage[colorlinks=true,citecolor=blue,linkcolor=blue,urlcolor=blue]{hyperref}

\begin{document}

\title{Large Topological Magnetic Optical Effects and Imaging of Antiferromagnetic Octupole Domains of  an Altermagnet-like Weyl Semimetal}

\author{Xingyue Han}
\affiliation{Department of Physics and Astronomy, University of Pennsylvania, Philadelphia, Pennsylvania 19104, USA}
\author{Xiaoran Liu}
\affiliation{Department of Physics and Astronomy, Rutgers, The State University of New Jersey, Piscataway, New Jersey 08854, USA}
\author{Mikhail Kareev}
\affiliation{Department of Physics and Astronomy, Rutgers, The State University of New Jersey, Piscataway, New Jersey 08854, USA}
\author{Jak Chakhalian}
\affiliation{Department of Physics and Astronomy, Rutgers, The State University of New Jersey, Piscataway, New Jersey 08854, USA}
\author{Liang Wu}
\email{liangwu@sas.upenn.edu}
\affiliation{Department of Physics and Astronomy, University of Pennsylvania, Philadelphia, Pennsylvania 19104, USA}

\date{\today}

\begin{abstract}

Pyrochlore iridates have attracted significant interest due to their complex phase behavior arising  from the interplay among electron correlations, quantum metric in flat bands, geometrically frustrated lattices, and topology induced by strong spin-orbit coupling. In this study, we focus on Eu$_2$Ir$_2$O$_7$ thin films oriented along the (111) crystallographic direction. This quantum material, identified as an antiferromagnetic Weyl semimetal, exhibits a large anomalous Hall effect in transport experiments. Here we employ optical circular dichroism microscopy, to directly image ferroic octupole order and resolve all-in--all-out and all-out--all-in antiferromagnetic domains below the Néel temperature. Remarkably, despite the absence of a detectable net magnetic moment at zero applied magnetic field, we detected a large magnetic circular dichroism signal ($\sim 10^{-4}$) and Kerr effect ($\sim 10^{-4}$ radians) in zero magnetic field attributable to Berry curvature effects from Weyl nodes. Eu$_2$Ir$_2$O$_7$ is a non-collinear magnet with vanishing net moment and magnetic octupole order, similar to the recently proposed collinear d-wave altermagnets, allowing for magneto-optical responses and anomalous Hall effect. This finding likely represents the first demonstration of magnetic circular dichroism and Kerr effect in a topologically non-trivial quantum antiferromagnet with a vanishing net magnetization. Our work opens up the possibility of ultrafast domain switching in the terahertz frequency and the domain wall dynamics in the magnetic Weyl systems, which establishes the foundation for topological antiferromagnetic spintronics.

\end{abstract}

\maketitle

\textbf{Introduction}

Research interest in pyrochlore iridates has surged since the theoretical prediction of novel topological states including the magnetic Weyl semimetal (WSM), axion insulator, and Luttinger metal \cite{Wan2011PRB, Yang2014PRL,kondo2015quadratic}. These materials, a family of 5$d$ iridium-based systems, exhibit strong spin-orbit coupling, geometrically frustrated lattice, and electron correlations, resulting in a rich electronic phase diagram \cite{ Wan2011PRB, nakatsuji2006metallic, machida2010time, Ueda2016PRB, tian2016field, Witczak2012PRB, Matsuhira2007JPSJ, Ueda2017NatComm, Ueda2014PRB, Ueda2020PRB}. The WSM phase is characterized by linearly dispersing bulk bands that intersect at Weyl points \cite{Armitage2018RMP, Hasan2017AnnualReview, Wan2011PRB, Yan2017AnnualReviews, ni2021natcomm} and topologically protected surface states with open contours known as Fermi arcs \cite{Armitage2018RMP, Hasan2017AnnualReview, Wan2011PRB, Yan2017AnnualReviews}. The realization of a WSM state requires the breaking of either time-reversal symmetry (TRS) or inversion symmetry. In TRS-breaking systems, the net accumulation of Berry curvature in momentum space leads to an intrinsic anomalous Hall effect (AHE). This phenomenon has been observed in ferromagnetic WSMs such as Co$_3$Sn$_2$S$_2$ \cite{Liu2018NatPhy, Wang2018NatComm} and Co$_2$MnGa \cite{Manna2018PRX, Anastasios2019PRB, han2022prb}, and noncollinear antiferromagnetic WSM such as Mn$_3$Sn \cite{Nakatsuji2015Nat, suzuki2017prb, khadka2020sciadv}. In contrast, in systems where TRS is preserved but inversion symmetry is broken, the net Berry curvature vanishes, thus diminishing the intrinsic AHE \cite{Yan2017AnnualReviews}, but enables interesting nonlinear optical responses\cite{ni2021natcomm, ni2020linear,wu2017giant,ma2021topology}.

\begin{figure*}
\includegraphics[width=0.8\textwidth]{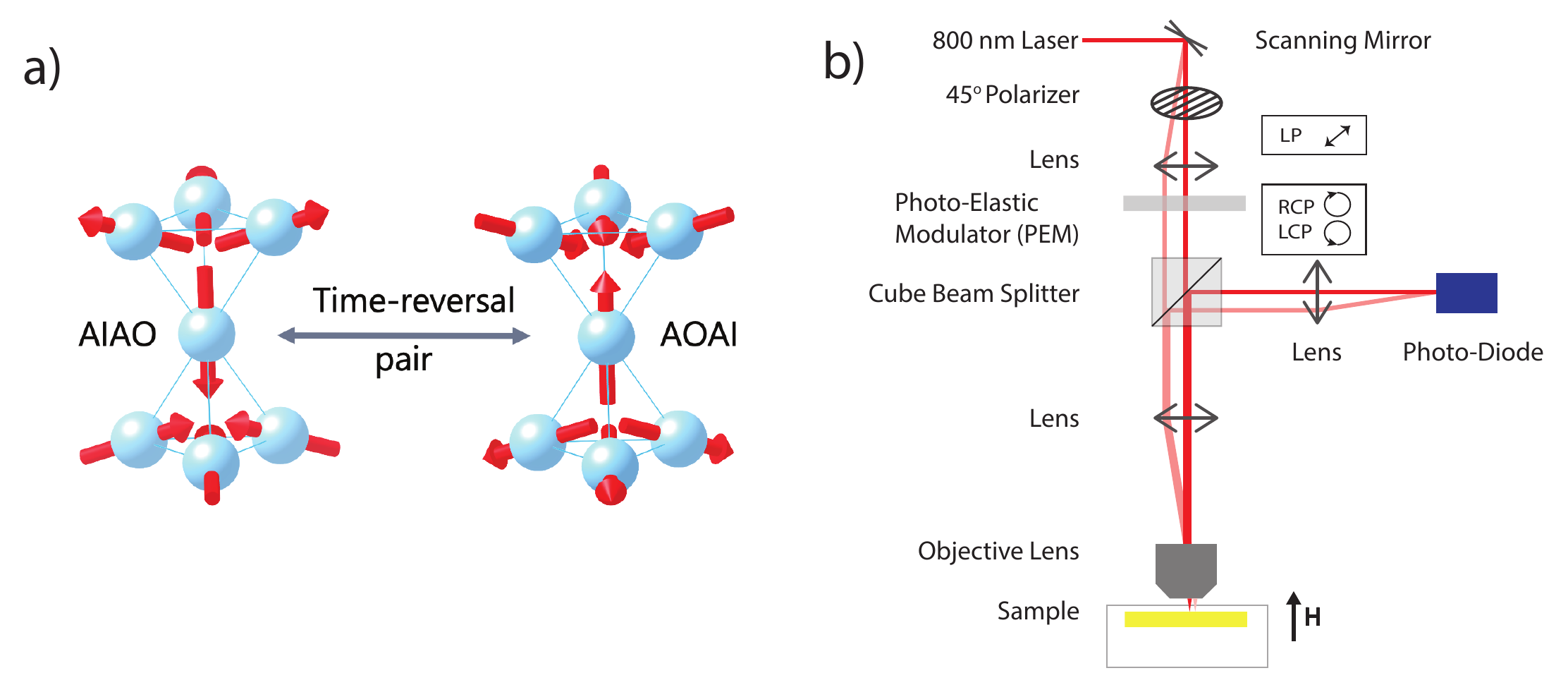}
\caption{\textbf{(a)} Schematic of the two spin configurations in pyrochlore iridates: all-in-all-out (AIAO) and all-out-all-in (AOAI). 
    \textbf{(b)} Optical setup for CD measurements. The system enables polarization modulation and raster scanning for imaging.  }
\label{Fig1}
\end{figure*}

Pyrochlore iridates, A$_2$Ir$_2$O$_7$ (where A represents Yttrium or lanthanides), crystallize in a face-centered cubic lattice with $Fd\bar{3}m$ symmetry. This structure consists of two networks of corner-sharing tetrahedra composed of A$^{3+}$ and Ir$^{4+}$ ions. Below the Néel temperature, the Ir$^{4+}$ spins adopt an all-in-all-out (AIAO) and all-out-all-in (AOAI) antiferromagnetic order, where all moments point either toward or outward from the center of tetrahedra \cite{zhao2011magnetic,Sagayama2013PRB, Arima2013JPSJ, Takatsu2014PRB, Tomiyasu2012JPSJ, Ueda2022PRB, Ueda2015PRL,wang2017prl, GL2025JAP} (see Fig.~\ref{Fig1}a).  The resulting magnetic order preserves vanishing magnetic dipole but sustains a finite magnetic octupole moment, therefore breaks TRS \cite{Arima2013JPSJ, liang2017natphy, wang2017prl}. The magnetic octupole moments can be regarded as a ferroic order that can be switched via field cooling \cite{Fujita2015SciRep, fujita2016prb, kozuka2017prb, Ma2015Science}. Theoretical calculations suggest that when the electron correlation strength of the Ir 5$d$ electrons falls within a specific range, the AIAO/AOAI order stabilizes the WSM phase \cite{Wan2011PRB, Liu2021PRL}. 


Among the A$_2$Ir$_2$O$_7$ family, Eu$_2$Ir$_2$O$_7$ has attracted particular interest due to experimental signatures of exotic topological phases. Optical conductivity measurements of Eu$_2$Ir$_2$O$_7$ single crystals reveal a linear low-frequency response, indicating the presence of three-dimensional linearly dispersing bands from Weyl semimetal state\cite{Sushkov2015PRB}. Recently, Eu$_2$Ir$_2$O$_7$ (111) thin films exhibit a large anomalous Hall effect\cite{Liu2021PRL}, but whether it is intrinsic still needs future studies of scaling relationship to the longitudinal resistivity. The AIAO/AOAI order in Eu$_2$Ir$_2$O$_7$ has been verified with a Néel temperature around 120 K by x-ray resonant magnetic scattering (XRMS) \cite{Sagayama2013PRB,Liu2021PRL} and muon spin rotation and relaxation ($\mu$SR) \cite{zhao2011prb}. However, direct imaging of magnetic domains has been challenging due to the small magnetic moment of Ir$^{4+}$ ($J_{eff}$=1/2) \cite{kozuka2017prb}, and the large absorption for neutron scattering \cite{klicpera2022prb}.  Scanning SQUID technique also fails to image domains in Eu$_2$Ir$_2$O$_7$ since Eu$^{3+}$ is nonmagnetic and the dipole moment of Ir$^{4+}$ is too small \cite{kozuka2017prb}. Finally, whether these magnetic Weyl semimetal manifest large magneto-optical responses have not been predicted due to the difficulty in density functional theory due to the strong correlation, and have yet not been studied experimentally.


In this work, we investigate the magnetic domain structure of Eu$_2$Ir$_2$O$_7$ (111) thin films using optical circular dichroism (CD) microscopy and Kerr effect. CD measures the differential reflectivity between left- and right-circularly polarized light. Both magneto-optical effects  serve as  probes of magnetic octupole order in Eu$_2$Ir$_2$O$_7$. We observe a large CD signal below the Néel temperature (approximately 100~K), and the sign of the CD depends on the polarity of an out-of-plane cooling field.  The CD and Kerr effect have a topological origin of the Berry curvature contribution. Our optical CD imaging also reveals the AIAO and AOAI domain structures across a millimeter-scale area in Eu$_2$Ir$_2$O$_7$ thin film for the first time \cite{Xu2022NatPhy}. Finally,  altermagnets have recently received increased attention for the prediction of an anomalous Hall effect and an associated magneto-optical Kerr effect in collinear AFMs with a vanishing net moment \cite{SmejkalPRX2022, SmejkalPRX2022_2}. Since the magnetic octupole order in Eu$_2$Ir$_2$O$_7$  has the same symmetry as a d-wave altermagnet, Eu$_2$Ir$_2$O$_7$  can be viewed as a non-collinear realization of collinear d-wave altermagnets, and thus allows for magneto-optical responses predicted for the latter\cite{oike2024nonlinear}. Our work demonstrates that optical CD and Kerr effect are  powerful probes for studying magnetic multipole materials with TRS breaking in magnetic Weyl semimetals and altermagnets. \\


\textbf{Large topological magnetic circular dichroism and Kerr effect}

We fabricate (111)-oriented Eu$_2$Ir$_2$O$_7$ thin films (40 nm and 100 nm) on (111) yttria-stabilized ZrO$_2$  substrates using pulsed laser deposition via the \textit{in situ} solid-phase epitaxy method \cite{Liu2021PRL, liu2020APL}. The (111) orientation is confirmed by high-resolution x-ray diffraction. The structural quality is examined using scanning transmission electron microscopy, while the presence of the AIAO magnetic ground state is demonstrated through x-ray resonant magnetic scattering.  Further details regarding sample characterization and transport properties can be found in previous reports \cite{Liu2021PRL}.


The optical CD microscopy setup is illustrated in Fig.~\ref{Fig1}b. This system enables CD measurements at a fixed sample position and across a area of a few hundred microns. We use 800 nm ultrafast laser with a repetition rate of 80 MHz. The penetration depth is larger than the film thickness \cite{Sushkov2015PRB}. As shown in Fig.~\ref{Fig1}b, the initial polarization is linearly polarized (LP) aligned to $45^\circ$ using a linear polarizer. The beam then passes through a photoelastic modulator (PEM), which introduces a phase retardation $\tau = \tau_0 \sin(\omega t)$ between the horizontal and vertical components. In our CD measurements, $\tau_0$ is set to $\pi/2$, modulating the polarization between right-circularly polarized (RCP) and left-circularly polarized (LCP) light at a frequency of $\omega = 42$ kHz\cite{Xu2022NatPhy}. The light is focused onto the sample using an objective lens, and the reflected light is collimated by the same lens before being collected by a photodiode. The sample is housed in a cryostat capable of reaching temperatures as low as 2 K and under an out-of-plane magnetic field up to 6 T. To generate CD maps, a scanning mirror is employed to raster the beam over a $100 \,\mu$m $\times$ $100 \,\mu$m area (illustrated by the pink light path in Fig.~\ref{Fig1}b). A set of lenses arranged in a 4$f$ geometry assists the imaging process, achieving a spatial resolution of 2 $\mu$m. \\

\begin{figure*}
\includegraphics[width=\textwidth]{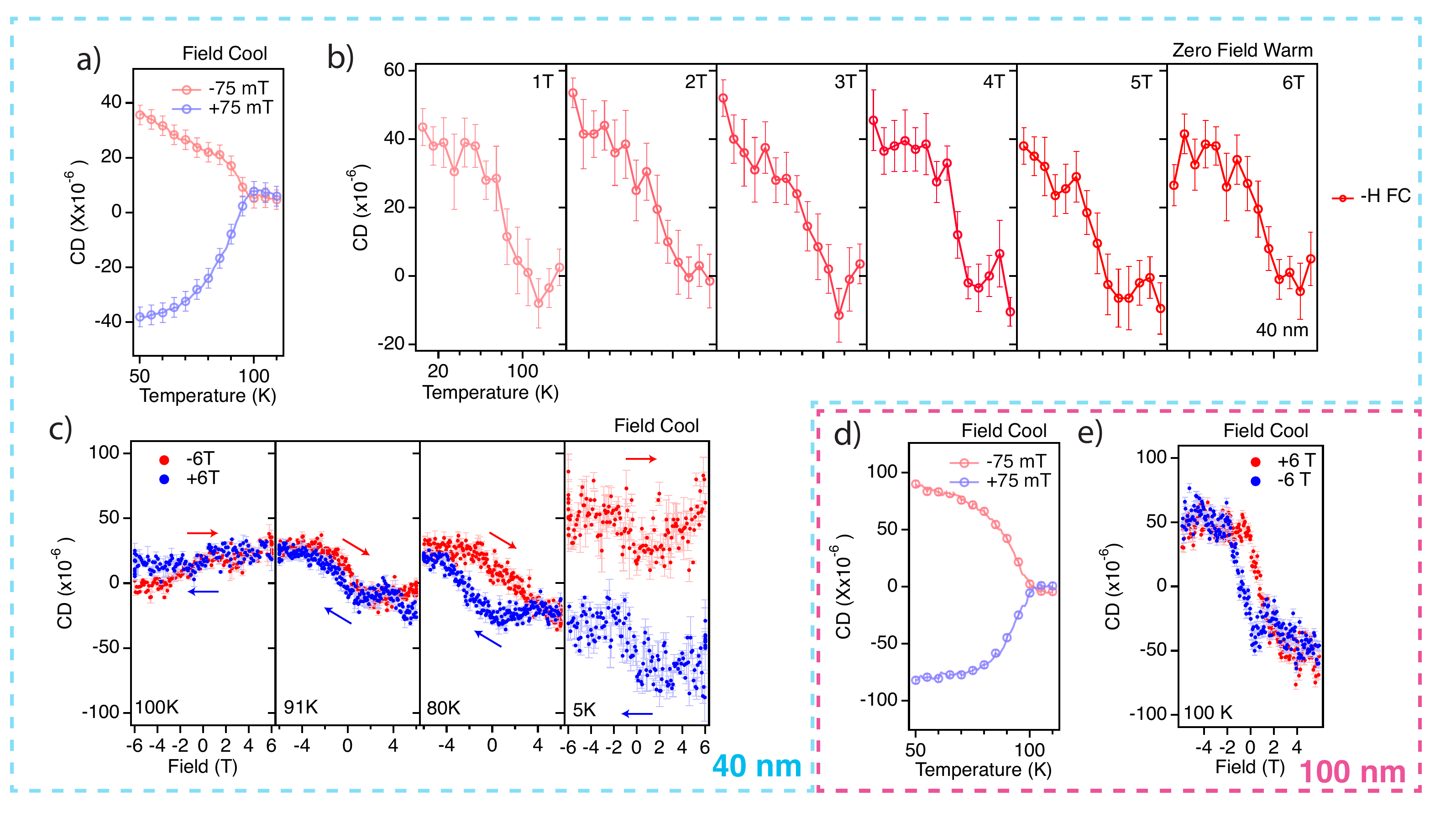}
\caption{
Circular dichroism (CD) measurements of Eu$_2$Ir$_2$O$_7$ thin films. 
\textbf{(a)} CD response of the 40 nm Eu$_2$Ir$_2$O$_7$ film under field cooling with $\pm$75 mT. 
\textbf{(b)} CD evolution during zero-field warming after field training under different field strengths. 
\textbf{(c)} CD response under field sweeps at various temperatures following $\pm$6 T field training. 
\textbf{(d)} CD response of the 100 nm Eu$_2$Ir$_2$O$_7$ film under field cooling with $\pm$75 mT. 
\textbf{(e)} CD response of the 100 nm film under field sweeps at 100 K following $\pm$6 T field training.
}

\label{Fig2}
\end{figure*}

We first train the 40 nm sample under an out-of-plane magnetic field of -75 mT during cooling. As shown in Fig.~\ref{Fig2}a, the abrupt onset of CD near 100 K indicates a magnetic phase transition. Reversing the magnetic field polarity switches the sign of the CD signal. The results in Fig.~\ref{Fig2}a are reproducible across multiple thermal cycles between 145 K and 50 K, confirming the stability and repeatability of the phase transition. We then perform field cooling under various magnetic field strengths ranging from -1 T to -6 T and record the CD signal during subsequent zero-field warming (Fig.~\ref{Fig2}b). The transition temperature shows no significant dependence on the cooling field, indicating the spontaneous TRS breaking. The CD consistently stabilizes into two states of similar amplitude and opposite sign at low temperatures. (We only show the negative field cooling results here.) Based on previous studies, the two states with opposite CD correspond to the AIAO and AOAI spin configurations \cite{Fujita2015SciRep}. The ability to switch between these domains by reversing the cooling field polarity suggests a ferroic nature of the magnetic octupole in Eu$_2$Ir$_2$O$_7$ \cite{Arima2013JPSJ}.


Next, we fix the temperature and investigate the CD response to varying magnetic fields. We train the 40 nm sample under $\pm$6 T from 140 K down to different temperatures. Fig.~\ref{Fig2}c highlights the results at 100 K, 91 K, 80 K, and 5 K. When field cooling stops near the phase transition (100 K), the CD versus field curve remains flat, characteristic of a paramagnetic state. When cooling stops slightly below the transition temperature (91 K), sweeping the magnetic field between -6 T and +6 T results in two stable states of opposite CD signs and similar amplitudes, corresponding to switching between AIAO and AOAI configurations. The hysteresis loop, with a coercive field of approximately 0.5 T at 91 K, becomes more pronounced at 80 K with a coercive field of 2 T. As the temperature decreases further, the coercivity continues to grow until it exceeds the maximum available field in our setup($\pm$6 T), resulting in an open loop at 5 K (Fig.~\ref{Fig2}c) \cite{Liu2021PRL, Fujita2015SciRep}. In this regime, sweeping the magnetic field within $\pm$6 T is insufficient to switch between the AIAO and AOAI configurations. Compared to transport measurements, both the phase transition temperature and coercive field observed in our CD results are lower. This discrepancy may arise from laser-induced heating, which locally elevates the temperature under ultrafast laser illumination. Similar behaviors are observed in the thicker 100 nm sample, as shown in Fig.~\ref{Fig2}d and Fig.~\ref{Fig2}e. The transition temperature is slightly higher than in the thinner sample, and the hysteresis loop adopts a more square-like shape, indicating stronger magnetic ordering in the thicker film.

\begin{figure}
\includegraphics[width=0.4\textwidth]{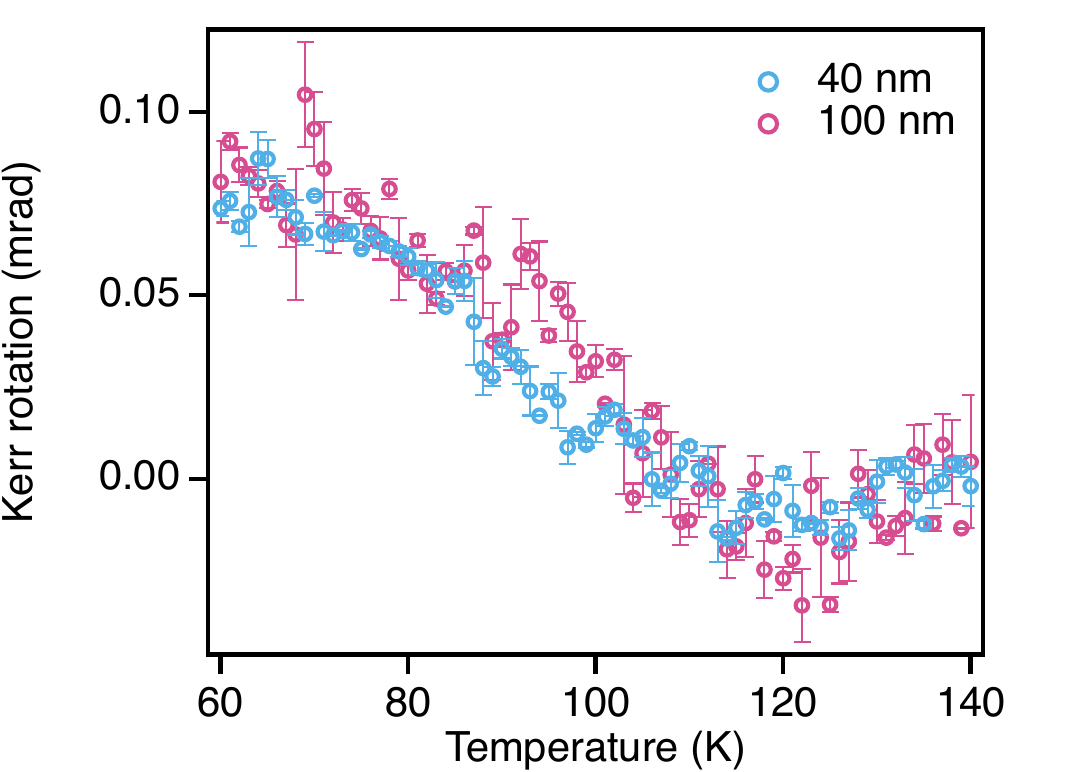}
\caption{
Kerr rotation of the 40 nm and 100 nm Eu$_2$Ir$_2$O$_7$ film under field cooling.
}

\label{Fig5}
\end{figure}

\begin{figure}
\includegraphics[width=0.5\textwidth]{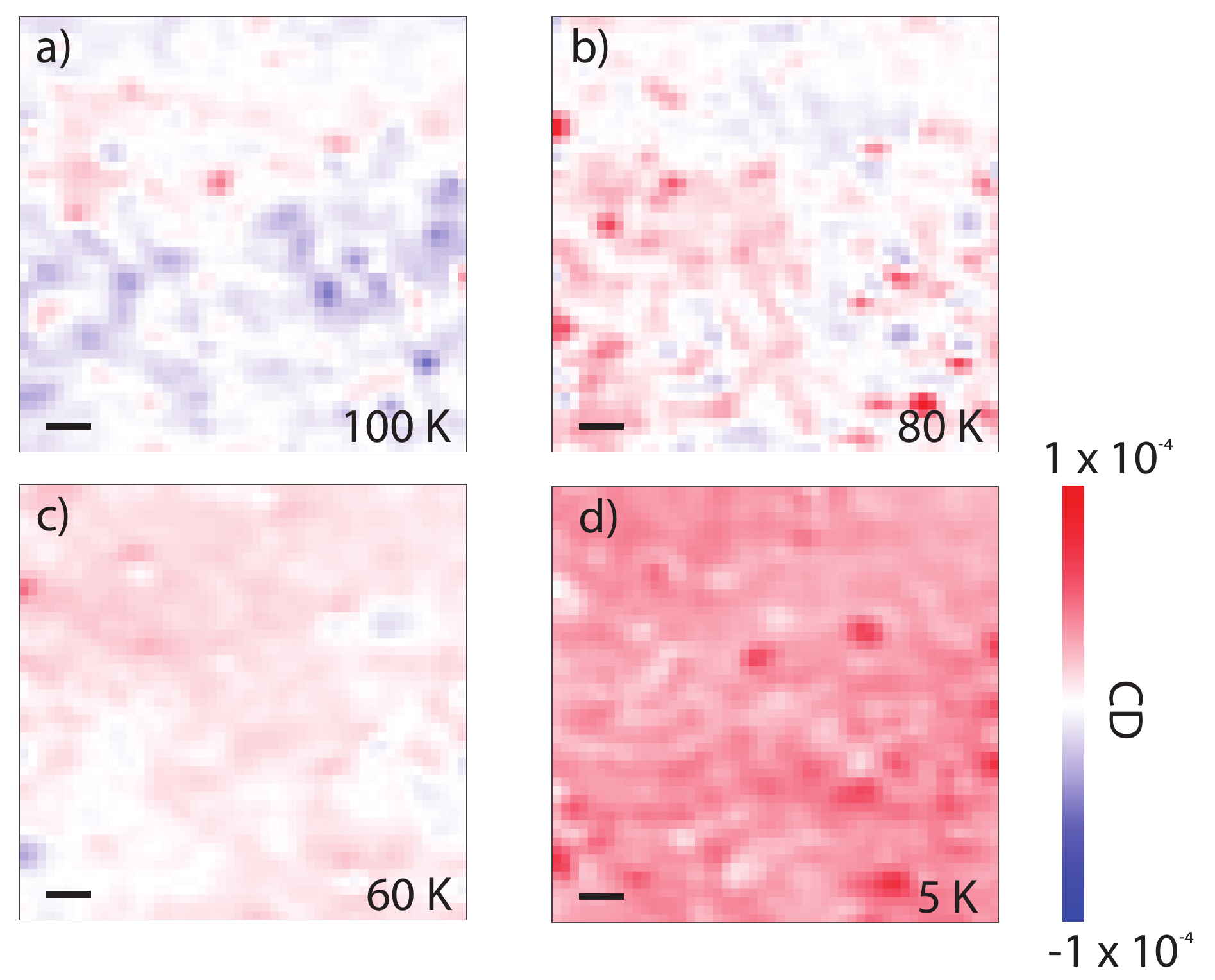}
\caption{
Circular dichroism (CD) maps of the 40 nm Eu$_2$Ir$_2$O$_7$ thin film under zero magnetic field at different temperatures: 
\textbf{(a)} 100 K, \textbf{(b)} 80 K, \textbf{(c)} 60 K, and \textbf{(d)} 5 K. 
The maps are obtained after field cooling under -6 T. Scale bars: 10 $\mu$m.
}

\label{Fig3}
\end{figure}

\begin{figure*}
\includegraphics[width=\textwidth]{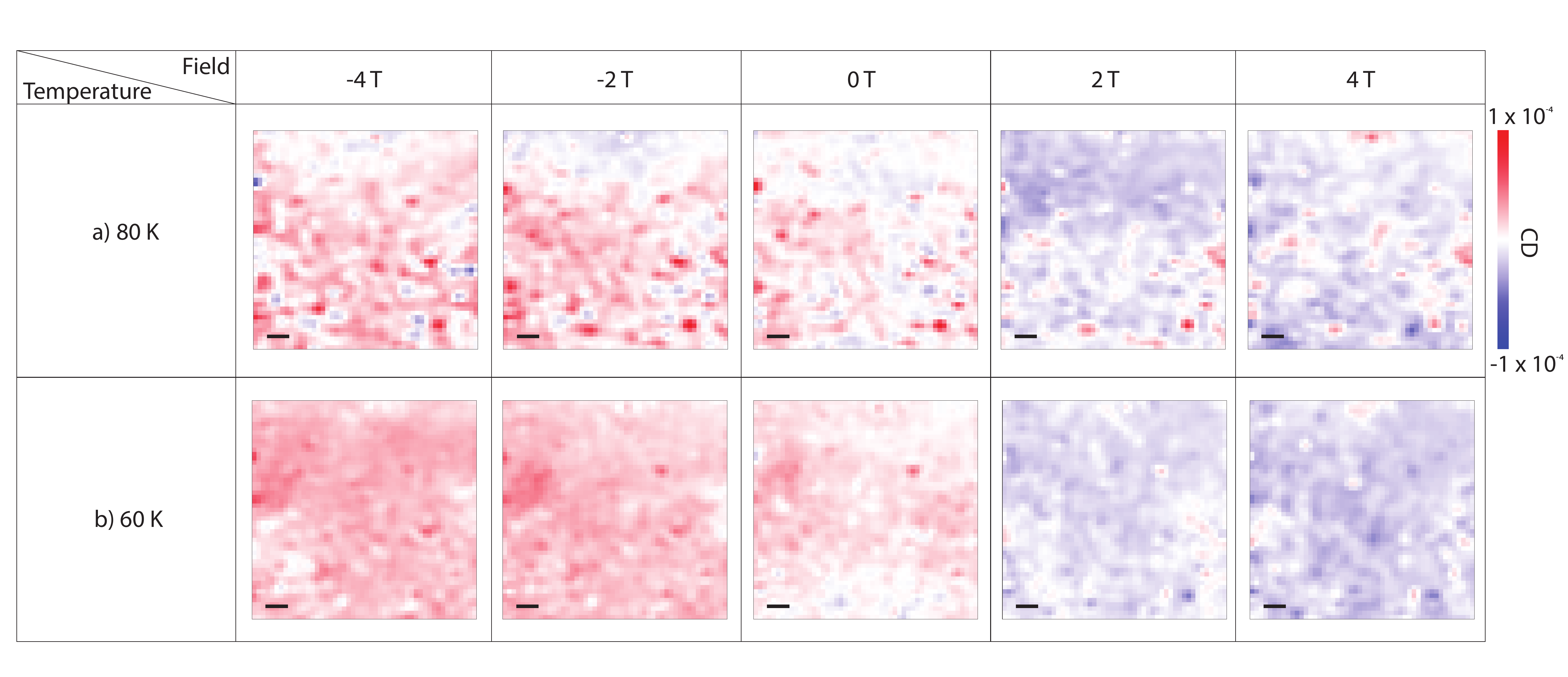}
\caption{
Circular dichroism (CD) maps of the 40 nm Eu$_2$Ir$_2$O$_7$ thin film during a field sweep at different temperatures: 
\textbf{(a)} 80 K and \textbf{(b)} 60 K. 
The maps are obtained after field cooling under -6 T. Scale bars: 10 $\mu$m.
}
\label{Fig4}
\end{figure*}

Since CD and Kerr effect obey the same symmetry requirements as the intrinsic AC anomalous Hall effect, it provides direct evidence of TRS breaking in Eu$_2$Ir$_2$O$_7$ \cite{Fumagalli2021, feng2015prb}:

\begin{equation}
\theta_K+i \epsilon_K=\frac{-\sigma_{x y}}{\sigma_{x x} \sqrt{1+i(4 \pi / \omega) \sigma_{x x}}}
\end{equation}


where $\theta_K$ is the Kerr angle, $\epsilon_K$ is the Kerr ellipticity, which is proportional to CD, $\sigma_{xx}$ is the longitudinal conductivity, and $\sigma_{xy}$ is the Hall conductivity. Here  $\theta_K$, $\epsilon_K$, $\sigma_{xx}$ and $\sigma_{xy}$ are at  the frequency of the light.  Kerr rotation and CD are related to each other by Kramers Kronig transformation. Thus, the reversal of the CD sign directly corresponds to a flip in the sign of $\sigma_{xy}$, reflecting the switching of the octupole moment between AIAO and AOAI states.  The temperature and field dependence of CD confirm the presence of two magnetic domains related by the time-reversal operator in Eu$_2$Ir$_2$O$_7$ thin films.  The advantage of CD and Kerr effect over DC anomalous Hall effect is that the latter also has a contribution from extrinsic effects such as side jump and skew scattering. Due to the high frequency in the optical measurements, the extrinsic effects are highly suppressed\cite{mazin2023altermagnetism}.  Therefore, we would to like to discuss the origin and the magnitude of the optical CD. We also measure the Kerr rotation by using a similar modulation technique with an PEM in these two  thin films during field cooling with $75$\,mT. As shown in Fig.~\ref{Fig5}, the Kerr angle exhibits a similar transition temperature of around 100 K. Previously, Kerr effect or CD have been observed in magneto-electric antiferromagnets (AFMs)\cite{huang2018electrical,jiang2018controlling,du2023topological}, canted AFMs\cite{keller2001magneto} and AFM Weyl semimetal Mn$_3$Sn\cite{higo2018natpho}, and they all have a net moment (see Fig.3 in Ref.\cite{higo2018natpho}). The peak of the Kerr rotation spectrum in Mn$_3$Sn is 3.5 $\times$ 10$^{-4}$ rad contributing from Berry curvature, but Mn$_3$Sn has a net moment around 0.002 $\mu$B/Mn\cite{higo2018natpho}. There is no detectable net moment In Eu$_2$Ir$_2$O$_7$ at 0 T, yet the CD and Kerr effect is also on the order of  10$^{-4}$ (see Figs.\ref{Fig2}-\ref{Fig4}).  Note that we might not have detected the peak of the spectrum, which might be even larger. Therefore, we conclude the large CD observed in Eu$_2$Ir$_2$O$_7$ contributes from Berry curvature and is topological in its origin. Considering the vanishing net moment, the magneto-optical responses in Eu$_2$Ir$_2$O$_7$ are more striking than the antiferromagnetic Weyl semimetal Mn$_3$Sn\cite{higo2018natpho}.   \\

\textbf{Optical imaging of antiferromagnetic octupole domains}

In Fig.~\ref{Fig3}, we present CD maps of the 40 nm Eu$_2$Ir$_2$O$_7$ sample under zero field at various temperatures (100 K, 80 K, 60 K, and 5 K) . The sample is first trained under -6 T, after which the external field is removed. In the maps, blue and red regions represent negative and positive CD, corresponding to the AIAO (or AOAI) and AOAI (or AIAO) domains, respectively. When field cooling stops at 100 K (Fig.~\ref{Fig3}a), near the transition temperature, the map exhibits weak remnant order with interspersed light red and blue regions. At 80 K (Fig.~\ref{Fig3}b), the red domain expands while the blue domain shrinks. By 60 K (Fig.~\ref{Fig3}c), the red domain dominates the area. At 5 K (Fig.~\ref{Fig3}d), the entire region is red, implying the formation of a single-domain state. The CD amplitude also increases with decreasing temperature, consistent with the temperature dependence shown in Fig.~\ref{Fig2}a.

Next, we investigate domain evolution in response to varying magnetic fields. Again, we train the sample under -6 T and then sweep the field while scanning the same $100 \,\mu$m $\times$ $100 \,\mu$m area in the sequence -4 T, -2 T, 0 T, 2 T, and 4 T. At 80 K (Fig.~\ref{Fig4}a), the sample starts in a multi-domain state. The negative cooling field favors the red domain, which is predominant at -4 T. As the field is increased to -2 T and 0 T, the blue domain gradually expands, although the red domain remains dominant, consistent with the hysteresis behavior in Fig.~\ref{Fig2}c. As the field approaches the coercive field at 2 T, most regions flip to the blue domain. At 4 T, although the blue domain dominates, the region remains in a multi-domain state. At a lower temperature of 60 K (Fig.~\ref{Fig4}b), domain evolution differs. The sample starts in a single-domain state at -4 T, which remains largely unchanged at -2 T. At 0 T, the red domain still dominates, though with reduced CD amplitude. Due to the large field step from 0 T to 2 T, domain wall propagation responsible for CD sign reversal is not observed. The maps at 2 T and 4 T exhibit good uniformity, suggesting a single-domain state with the opposite polarity.\\

\textbf{Discussion}

 CD in reflection, the imaginary part of the complex magneto-optical Kerr effect, has traditionally been associated with net magnetization in ferromagnetic and ferrimagnetic systems \cite{erskine1973prl, erskine1973prb}. Recently, a large MOKE was discovered in the noncollinear antiferromagnet Mn$_3$Sn \cite{higo2018natpho}, where it was attributed to cluster multipoles rather than a simple magnetic dipole \cite{feng2016prb, suzuki2017prb}. Similarly, in Eu$_2$Ir$_2$O$_7$, the magnetic octupole moment induces a nonzero Berry curvature \cite{Liu2021PRL}, accounting for the observed CD and Kerr rotation. Our results demonstrate that the octupole moment forms a ferroic order, determined by the polarity of the cooling field.

The CD maps provide the first direct visualization of domain structures originating from the Ir$^{4+}$ lattice in Eu$_2$Ir$_2$O$_7$ (111) thin films. We resolve AIAO and AOAI domains exhibiting opposite CD signs.  This advancement paves the way for ultrafast domain switching and domain dynamics and their potential applications for the topological antiferromagnetic spintronics.   To directly observe domain wall nucleation and propagation, finer field steps near the coercive field are required. The domain walls, serving as interfaces between potential Weyl semimetal bulk states, are predicted to host robust in-gap states \cite{GL2025JAP, Wan2011PRB}, but are beyond our spatial resolution. We hope future microcopy by scanning nitrogen vacancy center and nano-SQUID  can shed light on the exotic domain walls. Looking forward, we hope our work will generate interests in measuring CD and Kerr spectrum to search for larger magneto-optical effects and theoretical studies of magneto-optical properties in correlated Weyl semimetal to investigate whether the large magneto-optical response is due to intrinsic properties of the electronic bands encoded\cite{oike2024nonlinear}. Our work establishes optical CD and Kerr effect as  powerful probes for studying magnetic multipole materials with ferroic order in magnetic Weyl semimetals and altermagnets.\\

\textbf{Data and materials availability:} All data needed to evaluate the conclusions are present in the paper. Additional data related to this paper may be requested from the authors.\\

\textbf{Acknowledgements}
 X.H. and L.W. acknowledge the support by Air Force Office of Scientific Research under award no. FA955022-1-0410. The development of the scanning imaging microscope was partially sponsored by the Army Research Office and was accomplished under Grant Number W911NF-21-1-0131, W911NF-20-2-0166 and W911NF-25-2-0016. J. C. acknowledges the support by the U.S. Department of Energy, Office of Science, Office of Basic Energy Sciences under Award No. DE-SC0022160. X. L. acknowledges the Gordon and Betty Moore Foundation EPiQS Initiative through Grant No. GBMF4534.\\

\textbf{Author Contributions:} L.W. conceived and supervised the project.  X.H. performed the CD experiments and analyzed the data with L.W.. X.L., M.K. and J.C. grew the thin films. X.H. and L.W. wrote the manuscript from contributions of all authors. 

\textbf{Competing interests:} The authors declare that they have no competing interests.

\bibliography{EIO_NanoLetters}

\end{document}